\documentclass[useAMS,usenatbib]{mn2e}
\usepackage{graphicx,longtable,color}
\usepackage{mathrsfs}
\usepackage{epsfig}
\usepackage{natbib}
\usepackage{color}
\usepackage{float}
\usepackage{amsmath}
\usepackage{times}
\usepackage{upgreek}
\usepackage[varg]{txfonts}
\usepackage{enumerate}
\usepackage{verbatim}
\usepackage{booktabs}
\usepackage{multirow}
\usepackage{amssymb}
\usepackage{placeins}
\usepackage{bm}

\title[Alignment Probes The Origin of Cluster UDGs]{Exploring the origin of ultra-diffuse galaxies in clusters from their primordial alignment
}
\author[Rong et al.]{Yu Rong$^{1}$\thanks{E-mail: rongyuastrophysics@gmail.com}, Pavel E. Mancera Pi\~na$^{2,3}$, Elmo Tempel$^{4}$, Thomas H. Puzia$^{1}$, Sven De Rijcke$^{5}$\\
$^{1}$Instituto de Astrof\'isica, Pontificia Universidad Cat\'olica de Chile, Av. Vicu\~na Mackenna 4860, 7820436 Macul, Santiago, Chile\\
$^{2}$Kapteyn Astronomical Institute, University of Groningen, Landleven 12, 9747 AD, Groningen, The Netherlands\\
$^{3}$ASTRON, the Netherlands Institute for Radio Astronomy, Postbus 2, 7990 AA, Dwingeloo, The Netherlands\\
$^{4}$Tartu Observatory, University of Tartu, Observatooriumi 1, 61602 T\~oravere, Estonia\\
$^{5}$Ghent University, Dept. Physics \& Astronomy, Krijgslaan 281, S9, B-9000, Ghent, Belgium
}

\begin{document}
\maketitle

\begin{abstract}
We find that the minor axes of the ultra-diffuse galaxies (UDGs) in Abell~2634 tend to be aligned with the major axis of the central dominant galaxy, at a $\gtrsim 95\%$ confidence level. This alignment is produced by the bright UDGs with the absolute magnitudes $M_r<-15.3$~mag, and outer-region UDGs with $R>0.5R_{200}$. The alignment signal implies that these bright, outer-region UDGs are very likely to acquire their angular momenta from the vortices around the large-scale filament before they were accreted into A2634, and form their extended stellar bodies outside of the cluster; in this scenario, the orientations of their primordial angular momenta, which are roughly shown by their minor axes on the images, should tend to be parallel to the elongation of the large-scale filament. When these UDGs fell into the unrelaxed cluster A2634 along the filament, they could still preserve their primordial alignment signal before violent relaxation and encounters. These bright, outer-region UDGs in A2634 are very unlikely to be the descendants of the high-surface-brightness dwarf progenitors under tidal interactions with the central dominant galaxy in the cluster environment. Our results indicate that the primordial alignment could be a useful probe of the origin of UDGs in large-scale structures.
\end{abstract}
\begin{keywords}
galaxies: clusters: individual (A2634) \--- galaxies: statistics \--- galaxies: structure \--- galaxies: kinematics and dynamics
\end{keywords}
\section{Introduction}

The formation of ultra-diffuse galaxies \citep[UDGs;][]{vanDokkum15}, with large effective radii ($r_{\rm{e}}> 1.5\ \rm{kpc}$) and low surface brightness (e.g., the $r$-band mean effective surface brightness of $\langle\mu_{\rm{e}}(r)\rangle> 24\ \rm{mag/arcsec^2}$), is under debate. 
There are four prevailing formation models for UDGs. The first model proposes that UDGs are failed $L^*$ galaxies originated in the massive halos of $\sim 10^{12}\ M_{\odot}$, which is supported by the simulation of \cite{Yozin15} and high stellar velocity dispersion in DF~44 \citep{vanDokkum16}. The second kind of models propose that the UDGs are produced by a high dark matter- or stellar-specific angular momentum, which is favored by semi-analytic galaxy formation models \citep[e.g.,][]{Amorisco16,Rong17a} and observations of isolated UDGs \citep{Pina20}. The third model is that UDGs are originated from stellar feedback, which is also supported by hydrodynamical simulations \citep{DiCintio17,Martin19,Jiang19} and stellar population analysis based on spectroscopic data for a handful of UDGs \citep{Martin-Navarro19,Chilingarian19,Rong20}. Another possibility is that UDGs are the descendants of typical dwarf galaxies under tidal interactions with massive galaxies in galaxy clusters \citep[e.g.,][]{Carleton19,Venhola17}, which is favored by the studies of intrinsic morphology dependence on luminosity/environment \citep{Rong19a} and stellar orbits for several UDGs in galaxy clusters/groups \citep{Chilingarian19,Penny08}, as well as observed tidal features around UDGs \citep{Greco18,Bennet18}. In some simulations, UDGs are the products of mixture of the two or three models \citep[e.g.,][]{Jiang19,Liao19,Sales19}.

Galaxy alignments provide clues to the formation and evolution of galaxies. 
The studies of their primordial (direct) alignment, i.e., the orientations of elliptical galaxies or spin axes of spiral galaxies exhibit a tendency of alignment with the elongation of the host/closest galaxy cluster/large-scale filament, contributes to the understanding of the previous merger processes of galaxies, and particularly, the origin of angular momenta of galaxies \citep[e.g.,][]{Tempel15,Pahwa16,Rong15b,Rong16,Zhang15,Codis12,Libeskind13}. Radial alignment, i.e., the major axes of satellite galaxies are preferentially oriented toward the center of the parent cluster, however, is crucial to study the tidal interaction between the satellites and central dominant galaxy/parent cluster after galaxies falling into a cluster knot along the large-scale filaments, as well as mass distribution and dynamical state of the parent cluster \citep[e.g.,][]{Pereira08,Rong15a,Rong19b,Schneider13}.

The UDGs in the Coma and Abell~1314 clusters show the radial alignment signals \citep{Yagi16,Pina19}, suggesting that the member UDGs in clusters may be the products of strong tidal stripping or remarkably affected by tides. However, UDGs in the other galaxy clusters, including Fornax \citep{Venhola17,Rong19b}, 15 nearby clusters studied in \cite{Pina19} and \cite{vanderBurg16}, as well as two intermediate-redshift clusters A2744 \& AS1063 \citep[$z\sim 0.3$;][]{Lee17}, show no radial alignment signals, implying that the orientations of most UDGs in clusters are not strongly affected by the environment. Yet the possible primordial alignment of cluster UDGs, which may reveal their origin and evolution, has never been studied to the present. In this letter, we report the serendipitously discovered primordial alignment signal of UDGs in Abell~2634 galaxy cluster, i.e., the minor axes of UDGs tend to be aligned with the major axis of the central dominant galaxy of Abell~2634, which provides an important clue to the origin of the cluster UDGs. 

\section{Primordial Alignment of UDGs in A2634}

\begin{figure}
\centering
\includegraphics[width=0.8\columnwidth]{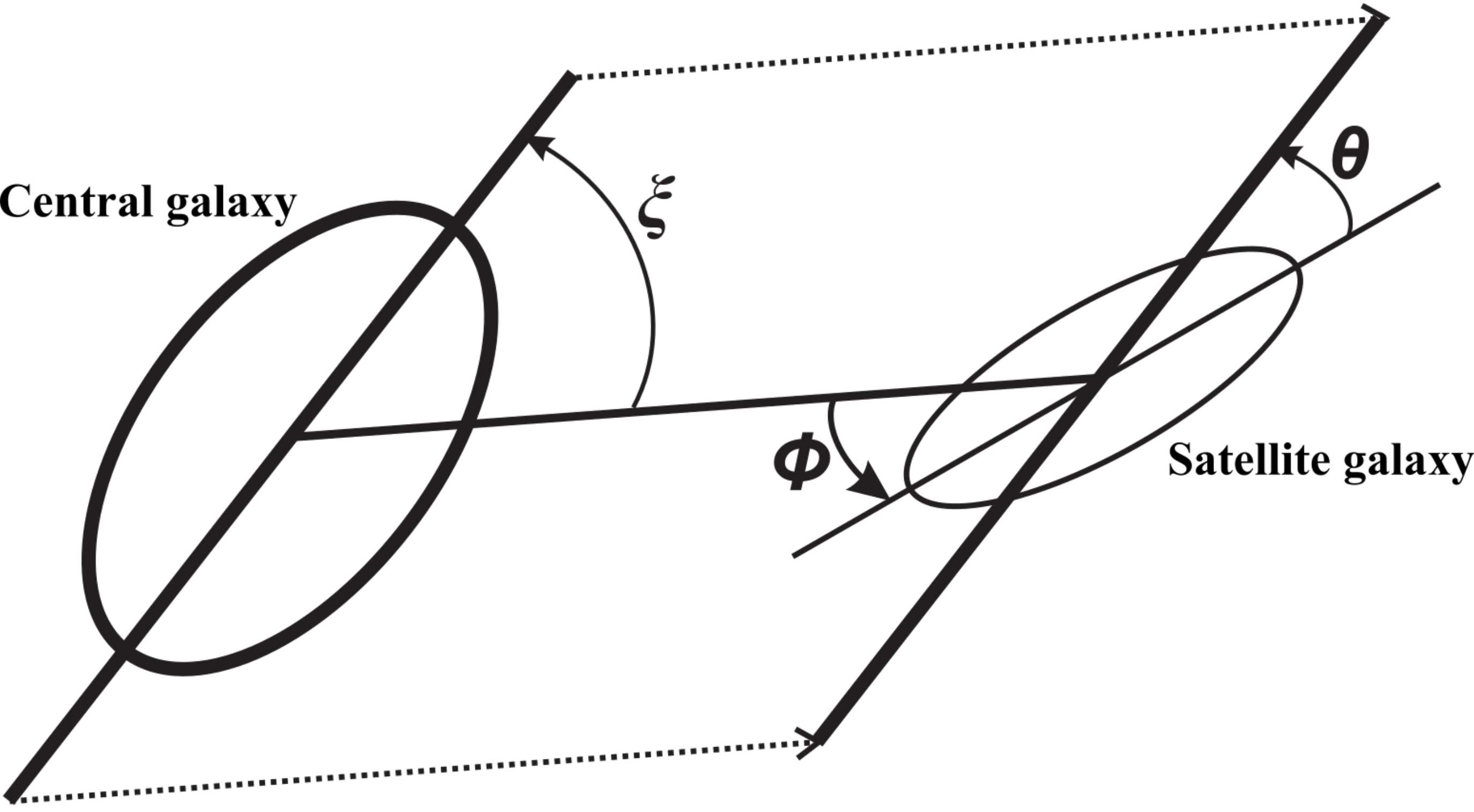}
\caption{The illustrations of the three angles $\theta$, $\phi$, and $\xi$, which are used to test for the primordial alignment, and radial alignment, as well as distributions of the satellite galaxies, respectively.}
\label{ske}
\end{figure} 

\begin{figure}
\centering
\includegraphics[width=\columnwidth]{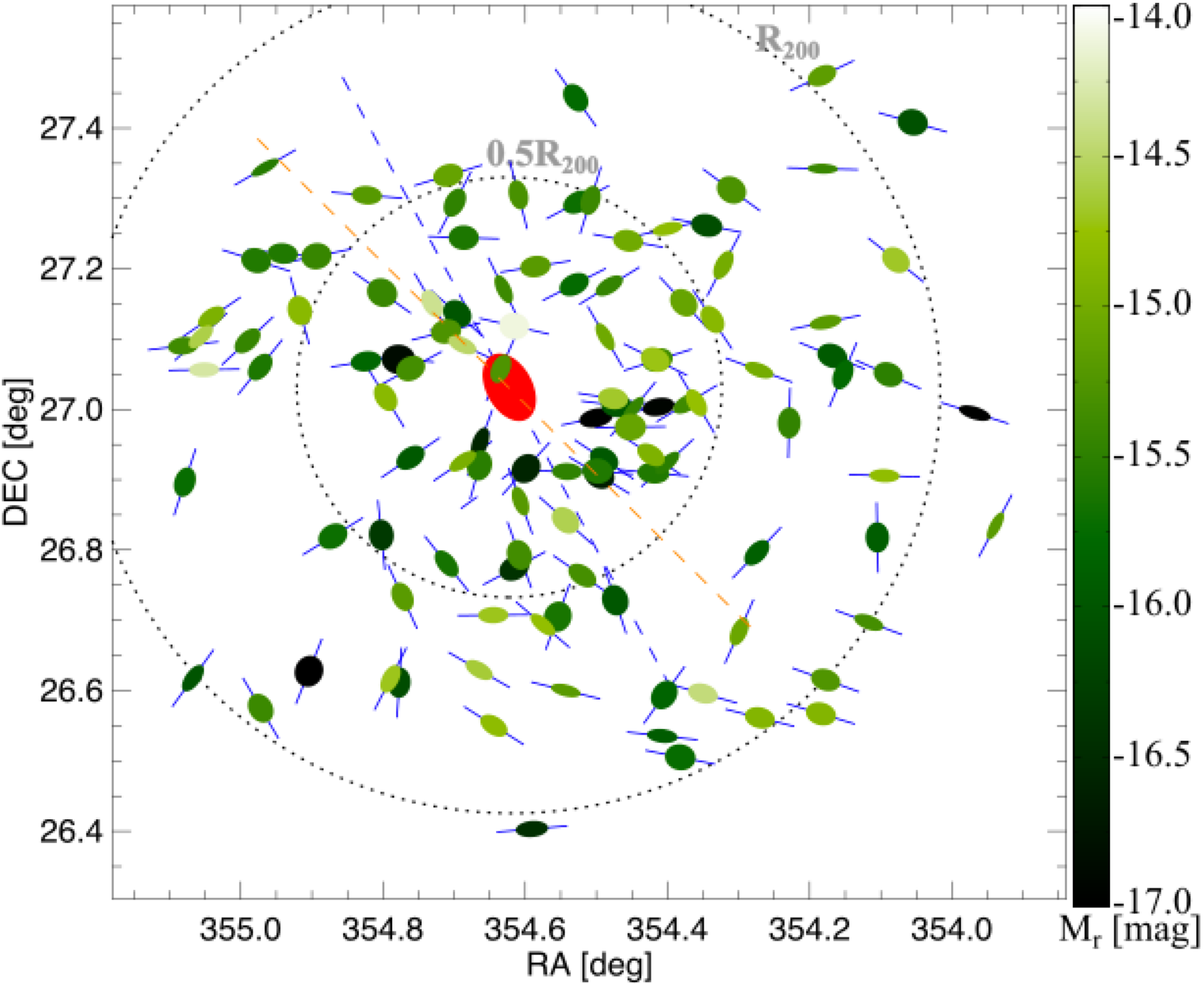}
\caption{The spatial distribution of UDGs in A2634. The dotted circles show the virial radius $R_{200}$ and $0.5R_{200}$. The cD galaxy NGC~7720 is highlighted with the red ellipse; the ellipses with the dark-green to light-green colors (the different colors reveal the different magnitudes as shown by the color bar) denote the UDGs (we only plot the ones with the projected axis ratios $b/a<0.9$). The blue solid and dashed lines show the major axes of the UDGs and NGC~7720, respectively; the orange dashed line shows the orientation of the elongation of X-ray emission.}
\label{dis}
\end{figure}

We use the UDG sample of \cite{Pina19} without performing any background subtraction, located in eight nearby clusters; the selection of UDGs is described in \cite{Pina19}. In this work, we only study the member UDGs with axis ratios of $b/a<0.9$, since the position angle uncertainties of the round UDGs with $b/a\geq 0.9$ are large. 
 
\begin{figure*}
\centering
\includegraphics[width=\textwidth]{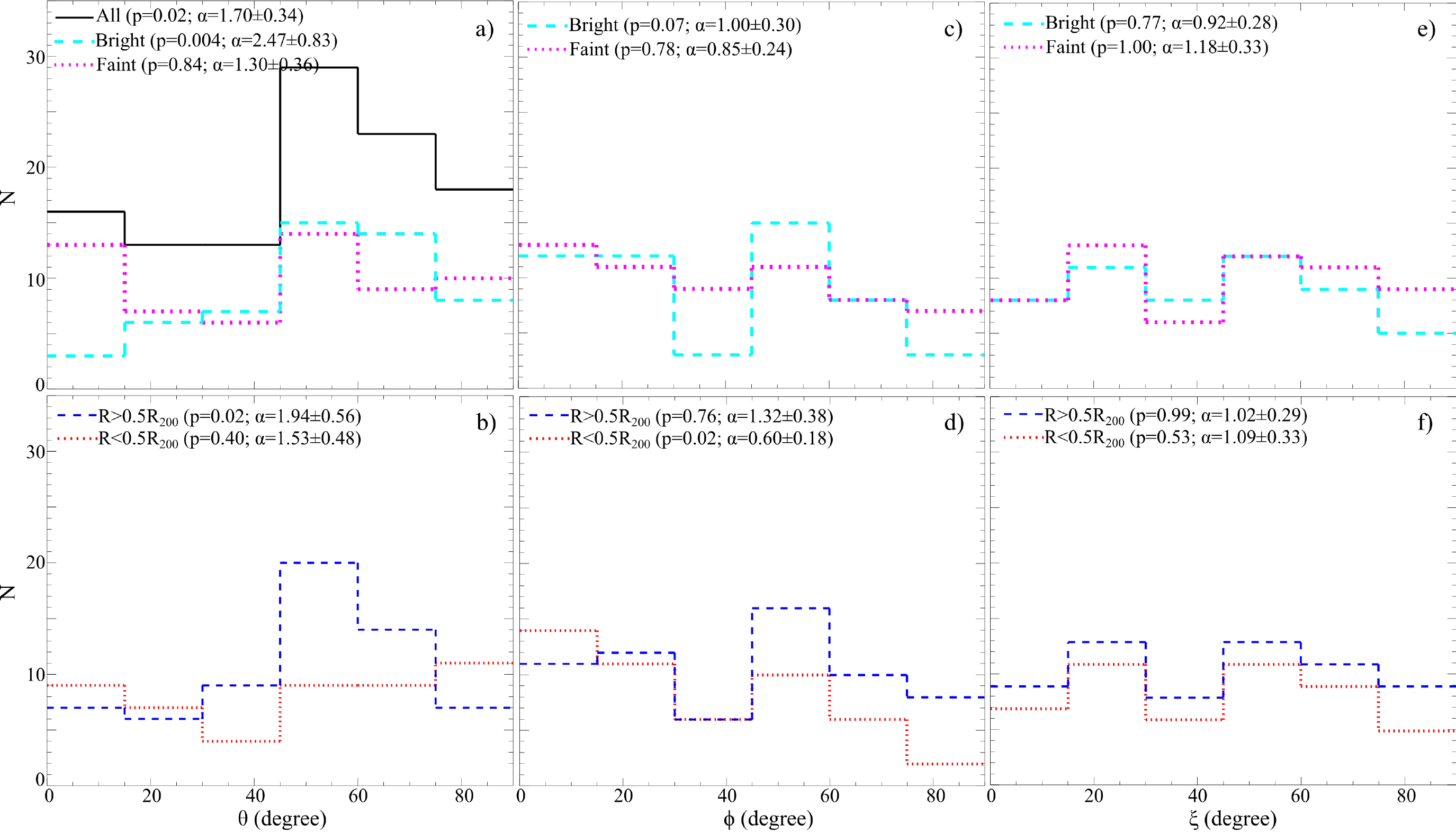}
\caption{Left panels: the PADs of the A2634 UDGs; Middle panels: the RADs of the A2634 UDGs; Right panels: the distributions of $\xi$ of the A2634 UDGs. The upper panels always show the distributions of the bright UDGs with $M_r<-15.3$~mag (dashed), and faint UDGs with $M_r>-15.3$~mag (dotted), respectively; the lower panels always show the distributions of the outer-region $R>0.5R_{200}$ (dashed) and inner-region $R<0.5R_{200}$ (dotted) UDGs, respectively. In panel~a, the PAD for all of the UDGs in A2634 is also shown by the solid histogram. In each panel, the $p$ value from the K-S test of comparing each distribution with a uniform distribution, as well as $\alpha$ value denoting significance of the excess of $>45^{\circ}$, are also shown.}
\label{PA_dis}
\end{figure*}

We investigate the distribution of position angles (PADs) of UDGs in each cluster; the position angle $\theta$ (in a range of $0\--90^{\circ}$), as shown in Fig.~\ref{ske}, is defined as the intersection angle between the major axes of a satellite UDG and the central dominant (cD) galaxy. The Kolmogorov-Smirnov (K-S) tests (i.e., comparing PAD with a uniform distribution) indicate that only the UDGs in A2634 (for A2634, there are 120 UDGs in the original sample, among which 112 UDGs have $b/a<0.9$, as shown in Fig.~\ref{dis}) present a significant alignment signal ($p\sim 0.02$), as shown in panel~a of Fig.~\ref{PA_dis}.

We find that the major axes of the UDGs in A2634 tend to be perpendicular to the major axis of the cD galaxy, i.e., $\theta$ are more likely to be $\theta>45^{\circ}$ rather than $\theta<45^{\circ}$. Analogous to the work of \cite{Yagi16}, we utilize a ratio $\alpha$ between the numbers of galaxies with $\theta>45^{\circ}$ and $\theta<45^{\circ}$, i.e., $\alpha\equiv N_{1}/N_2$ (where $N_1$ and $N_2$ are the UDG numbers of $\theta>45^{\circ}$ and $\theta<45^{\circ}$, respectively), to quantitatively{\footnote{The uncertainties of $\alpha$ are estimated with the bootstrap method.}} assess the significance of the excess of $\theta>45^{\circ}$ (i.e., more likely to be perpendicular rather than parallel to the major axis of the cD galaxy) in the PAD (a uniform distribution corresponding to $\alpha \simeq 1$, while a perpendicular alignment corresponding to $\alpha\gg1$). We find $\alpha=1.70\pm0.34$ for the PAD of the member UDGs in A2634, suggesting that there is an $\theta>45^{\circ}$ excess of the PAD at $\sim 2\sigma$ confidence level{\footnote{The significance of the excess is estimated by $(\alpha-1)/$err$_{\alpha}$, where err$_{\alpha}$ is the uncertainty of $\alpha$.}}.

The alignment signal in the PAD of UDGs in A2634 is primarily produced by the brighter/higher-mass UDGs with $M_r<-15.3$~mag (-15.3~mag is the median absolute magnitude of the UDGs in A2634), as shown in panel~a of Fig.~\ref{PA_dis}. The bright UDGs present a K-S test $p\sim 0.004$ and $\alpha \simeq 2.47\pm 0.83$. In contrast, the faint/low-mass UDGs with $M_r>-15.3$~mag exhibit no alignment signal with $p\sim 0.84$ and $\alpha \simeq 1.30\pm 0.36$.

\begin{figure*}
\centering
\includegraphics[width=0.8\textwidth]{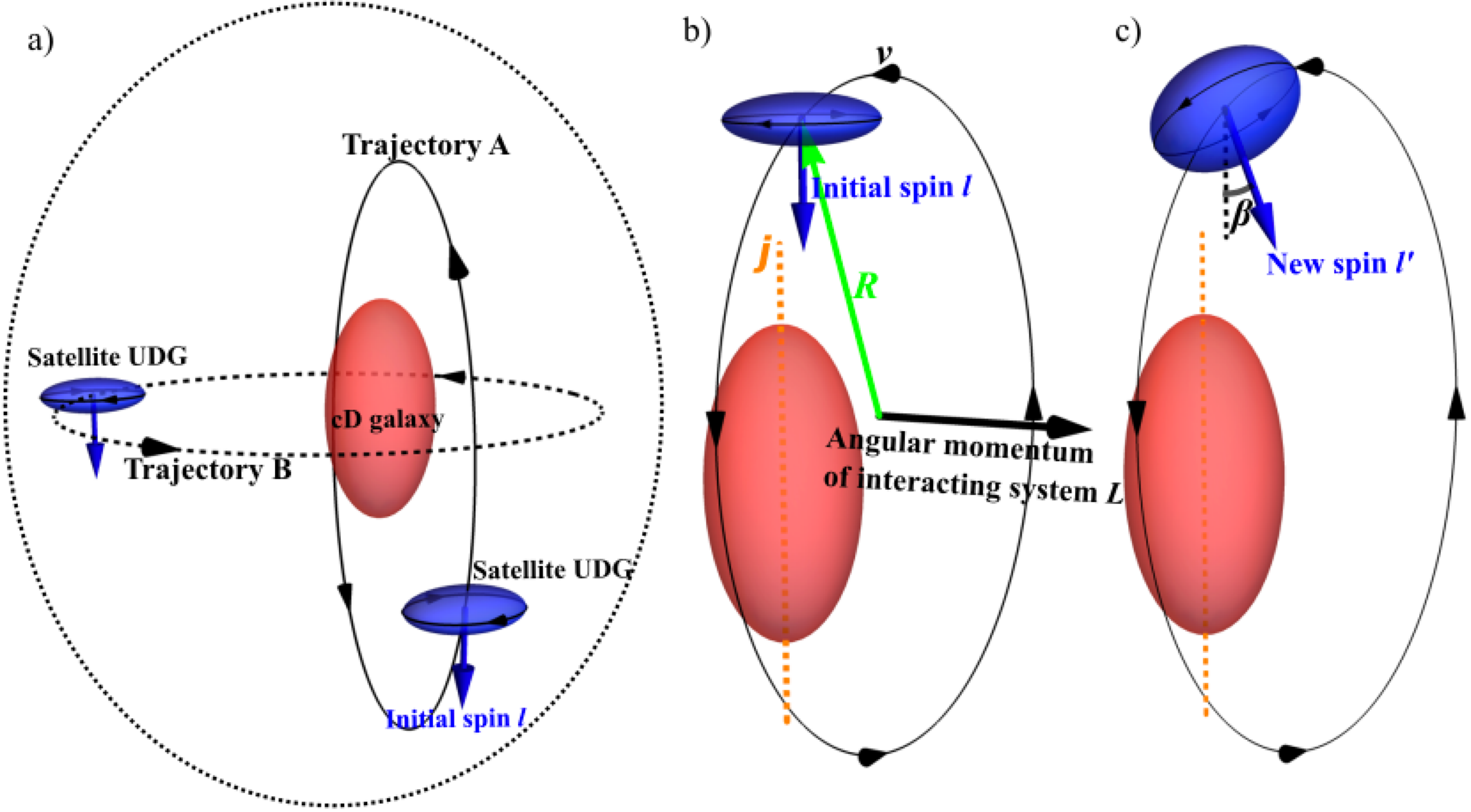}
\caption{Panel~a: the sketch of a satellite UDG motion in A2634. The big dotted ellipse denotes the A2634 cluster, and the red and blue ellipsoids show the cD galaxy and satellite UDG, respectively; the primordial spin {\bm{$l$}} of the satellite UDG is also shown by the blue arrows. We explore the two possible motion trajectories (A\& B) of the satellite UDG in the cluster. Panel~b: by assuming a motion trajectory~A, the angular monmentum of the interacting system (UDG and the cD galaxy) {\bm{$L$}}$\sim${\bm{$R$}}$\times${\bm{$v$}} tends to be perpendicular to the primordial spin {\bm{$l$}} of the satellite UDG and major axis {\bm{$j$}} (orange dashed line) of the cD galaxy. Panel~c: Affected by the the angular momentum of the interacting system, the new spin of the satellite UDG, {\bm{$l'$}}, has a intersection angle, $\beta$, with the major axis (dashed line) of the cD galaxy.}
\label{ske_L}
\end{figure*} 

The major axis of a cD galaxy of a galaxy cluster is found to strongly coincide with the orientation of its parent cluster and elongation of the host/closest large-scale filamentary structure \citep{Niederste-Ostholt10,West17,Fuller99,Struble90}. In A2634, the major axis of the cD galaxy, NGC~7720, shows a position angle of $\sim28^{\circ}$ \citep{Alam15}, and is also aligned to the elongation of X-ray emission and galaxy spatial distribution of A2634 \citep{Pinkney93,Scodeggio95}. Therefore, the results also suggest that the major axes of the bright UDGs in A2634 tend to be perpendicular to the elongation of the host cluster/large-scale filament, or that the minor axes, which coarsely show the (projected) spin axes of UDGs \citep{Zhang09,Mo98,Debattista15} since these cluster UDGs are very likely to have a thick-disky intrinsic morphology \citep{Rong19a}, are preferentially aligned with the elongation of the host cluster/large-scale filament. This preferential alignment resembles the spin alignment of spiral galaxies in dynamically-young clusters and large-scale filaments \citep{Tempel13a,Tempel13b}.

The preferential alignment of the bright UDGs in A2634 could be naturally explained as a spin alignment. Analogous to the spin acquisition of the low-mass spiral galaxies with halo masses $<5\times 10^{12}\ M_{\odot}$, the angular momentum of a UDG might be the result of the vorticity of the cosmic flow field on a scale greater than the virial radius, and thus be parallel to the elongation of the host/closest large-scale filament \citep[e.g.,][]{Codis12,Libeskind13,Laigle15,Veena19}. When these UDGs fell into the gravitational potential of the cluster along the large-scale filament, they can, to a certain extent, still preserve their primordial spin alignment before the violent relaxation and (impulsive) encounters with the cD galaxy \citep{Porciani02,Zhang15}. We note that A2634 is an unrelaxed cluster \citep{Pinkney93,Scodeggio95}; particularly, its member spiral galaxies exhibit both a multimodal velocity distribution and a much larger velocity dispersion than the member ellipticals, representing a dynamically young cluster population \citep{Scodeggio95}. 

Based on this model, we should expect that the UDGs in the outer region of cluster to show more significant preferential alignment signal, compared with those in the inner region, since the inner-region UDGs might be more easily affected by the tidal interactions with the cD galaxy, whereas the outer-region ones are less affected by tides and thus more likely to preserve their primordial alignment signal. As shown in panel~b of Fig.~\ref{PA_dis}, indeed, we find that the preferential alignment signal of the A2634 UDGs is primarily produced by the projected-outer-region (i.e., the cluster-centric distances $R>0.5R_{200}$, where $R_{200}$ is the virial radius $\sim 1.31\pm0.13$~Mpc) UDGs ($p\sim0.02$, $\alpha=1.94\pm0.56$), while the projected-inner-region UDGs ($R<0.5R_{200}$) exhibit no primordial alignment signal ($p\sim0.40$, $\alpha=1.53\pm0.48$).

In summary, the minor-axis alignment signal indicates that a large fraction of the member UDGs in A2634, particularly those with $M_r<-15.3$~mag and $R>0.5R_{200}$, are very likely to acquire their angular momenta from the vortices in the cosmic web; these UDGs then fell into the cluster along the large-scale filament, being weakly affected by the tides from the cD galaxy of the cluster, and thus could still preserve their primordial spin alignment (i.e., minor-axis alignment) signal before the violent relaxation and encounters. It also suggests that these bright, outer-region UDGs might form their extended stellar components outside of the cluster, since their morphologies/sizes should have not been much changed by the tides from the cD galaxy, i.e., they might be born as UDGs.

We also note that the tidal interaction models \citep[e.g.,][]{Carleton19} cannot reproduce such a minor-axis alignment of UDGs in a cluster environment.

\section{Discussion}

\subsection{Why the PAD excess does not peak at $\theta=90^{\circ}$?}

It is also worth to note that the PAD excess in Fig.~\ref{PA_dis} does not peak at $\theta=90^{\circ}$, but at $\theta\sim 50\--70^{\circ}$. We sketch a toy model which may be compatible with the biased PAD peak; it may be due to the interactions between UDGs and cD galaxy. As shown in Fig.~\ref{ske_L}, since the satellite UDGs (blue ellipsoids) with primodial spins {\bm{$l$}} (parallel to the major axis {\bm{$j$}} of the cD galaxy) should fall into A2634 along the elongation of the parent cluster/large-scale filament (i.e., {\bm{$l$}}//{\bm{$j$}}), a satellite UDG is more likely to orbit the cD galaxy by following the trajectory~A shown in panel~a of Fig.~\ref{ske_L}, rather than trajectory~B, i.e., the normal vector of the orbital plane {\bm{$n$}}$\perp${\bm{$j$}}. Therefore, if viewed in a co-moving reference frame with origin in the cD galaxy, during a fly-by, the angular momentum vector {\bm{$L$}}$\sim${\bm{$R$}}$\times${\bm{$v$}} (and {\bm{$L$}}//{\bm{$n$}}) of the interaction system, made up of the UDG and cD galaxy, should be perpendicular to {\bm{$l$}}, i.e., {\bm{$L$}}$\perp${\bm{$l$}}, as shown in panel~b of Fig.~\ref{ske_L}. As indicated in panel~c of Fig.~\ref{ske_L}, the primordial UDG angular momentum {\bm{$l$}} would be twisted by {\bm{$L$}} during the fly-by, and thus the new spin {\bm{$l'$}} of the UDG should have a intersection angle $\beta$ ($\beta\neq 0^{\circ}$) with the major axis {\bm{$j$}} of the cD galaxy, i.e., the major axes of the UDG and NGC~7720 should prefer to have a intersection angle $\theta=90^{\circ}-\beta$. The twisted angle $\beta$ depends on the interaction time and $\lvert${\bm{$L$}}$\rvert$. For the inner-region UDGs, which may pass through their pericenters at least once, the tidal interaction between the UDG and cD galaxy may be strong enough \citep[cf., e.g.,][]{Carleton19,Errani15} to significantly twist the primordial spins {\bm{$l$}}, and result the new spins {\bm{$l'$}} to be more likely to align with {\bm{$L$}} and the major axes of the inner-region UDGs to point towards the cD galaxy. Yet, for the outer-region UDGs which may have not pass through the pericenters, their primordial spins might be slightly affected and $\beta$ is also small. In panel~d of Fig.~\ref{PA_dis}, we also show the radial angle $\phi$ (cf. Fig.~\ref{ske}) distributions (RADs) of the inner-region and outer-region UDGs, and find that the inner-region UDGs exhibit a radial alignment signal ($p=0.02$, $\alpha=0.60\pm0.18$), probably suggesting that their primoridal spins (and orientations) have been completely twisted, while the outer-region ones have no radial alignment ($p=0.76$, $\alpha=1.32\pm0.38$) but show primordial alignment, plausibly suggesting the small $\beta\sim 20\--40^{\circ}$.

We also test whether the fact that the biased peak is attributed to the misalignment between the major axis of the cD galaxy and elongation of the host large-scale filament/X-ray emission \citep[PA$\simeq 45^{\circ}$; cf. also Fig.~\ref{dis};][]{Eilek84}. We find that the peak of the PAD excess would not be shifted towards $\theta=90^{\circ}$ by using the orientation of the X-ray emission elongation, and thus, it is not caused by the misalignment between the two axes. The projection effect of the large-scale filament (since the filament may have an intersection angle with the line-of-sight of the observer) also cannot explain the biased peak.

Note however, our results cannot exclude a possibility that the A2634 UDGs may have two different origins: one relatively-young population of bright UDGs located at the outer region, which acquired their spins from the vorticity of the cosmic flow field, and were accreted into the cluster later along the large-scale filament, as well as an old population of faint UDGs located primarily in the inner region, which were perhaps transformed from the high-surface-brightness typical dwarf progenitors under the tidal interactions with the cD galaxy in the cluster environment.

\subsection{General discussion}

The work of \cite{Rong15b} suggests that a fake primordial alignment signal may be the result of a combination of the radial alignment and anisotropic galaxy distribution (i.e., $\phi$ and $\xi$ shown in Fig.~\ref{ske} simultaneously tend to be 0). In panels~c, d and~e, f of Fig.~\ref{PA_dis}, we show the RADs and distributions of $\xi$ of the UDGs in A2634, respectively. The bright UDGs as well as outer-region UDGs exhibit isotropic spatial distributions, suggesting that the primordial alignment signals of the bright and outer-region UDG samples in panel~a \& b of Fig.~\ref{PA_dis} are `real'. We also note that the different $b/a$ cuts of UDG sample from 0.9 to 0.85, 0.8, 0.75, and 0.7 does not change our conclusion.

We only find a 2$\sigma$-confidence-level primordial alignment signal in A2634. However, note that the non-detected primordial alignment signals among the UDGs in the other seven clusters \cite{Pina19} may be primarily due to the fact that these clusters are much dynamically older than A2634 (e.g., A1314), and therefore, the primordial alignment of their member UDGs have been disrupted, as well as that A2634 is the richest cluster in the cluster sample of \cite{Pina19}, and the UDG numbers of the other clusters are too low to reveal the possibly existed alignment signals. Based on our model, we predict that researchers may find more spin alignment signals for UDGs in more galaxy clusters, particularly in the intermediate-redshift, dynamically-young clusters, where the member galaxies may still retain their primordial alignment signals \citep{Rong15b}.


\section*{Acknowledgments}

We thank the referee for their helpful comments, and X.-Y. Dong for her helps on plots. Y.R. thanks R. F. J. van der Burg, J. H. Lee, and M. G. Lee for their data support, and acknowledges the funding supports from FONDECYT Postdoctoral Fellowship Project No.~3190354 and NSFC grant No.\,11703037. P.E.M.P. is supported by the Netherlands Research School for Astronomy (Nederlandse Onderzoekschool voor Astronomie, NOVA), Phase-5 research programme Network 1, Project 10.1.5.6. E.T. was supported by ETAg grant IUT40-2 and by EU through the ERDF CoE grant TK133 and MOBTP86. T.H.P. acknowledges support through FONDECYT Regular project 1161817 and CONICYT project Basal AFB-170002.


\section*{Data Availability}

Data available on request.

\bibliographystyle{mn2e}

\begin{thebibliography}{}

\bibitem[\protect\citeauthoryear{Alam et al.}{2015}]{Alam15} Alam, S., et al., 2015, ApJS, 219, 12
\bibitem[\protect\citeauthoryear{Amorisco \& Loeb}{2016}]{Amorisco16} Amorisco, N. C., Loeb, A. 2016, MNRAS, 459, L51
\bibitem[\protect\citeauthoryear{Bennet et al.}{2018}]{Bennet18} Bennet, P., Sand, D. J., Zaritsky, D., Crnojevi\'c, D., Spekkens, K., Karunakaran, A. 2018, ApJ, 866L 11
\bibitem[\protect\citeauthoryear{Carleton et al.}{2019}]{Carleton19} Carleton, T., Errani, R., Cooper, M., Kaplinghat, M., Pe\~narrubia, J., Guo, Y. 2019, MNRAS, 485, 382
\bibitem[\protect\citeauthoryear{Chilingarian et al.}{2019}]{Chilingarian19} Chilingarian, I. V., Afanasiev, A. V., Grishin, K. A., Fabricant, D., Moran, S. 2019, ApJ, 884, 79
\bibitem[\protect\citeauthoryear{Codis et al.}{2012}]{Codis12} Codis, S., Pichon, C., Devriendt, J., Slyz, A., Pogosyan, D., Dubois, Y., Sousbie, T., 2012, MNRAS, 427, 3320
\bibitem[\protect\citeauthoryear{Debattista et al.}{2015}]{Debattista15} Debattista, V. P., van den Bosch, F. C., Roskar, R., Quinn, T., Moore, B., Cole, D. R. 2015, MNRAS, 452, 4094
\bibitem[\protect\citeauthoryear{Di Cintio et al.}{2017}]{DiCintio17} Di Cintio, A., Brook, C. B., Dutton, A. A., Macci\`o, A. V., Obreja, A., Dekel, A. 2017, MNRAS, 466L, 1
\bibitem[\protect\citeauthoryear{Eilek et al.}{1984}]{Eilek84} Eilek, J. A., Burns, J. O., O'Dea, C. P., Owen, F. N., 1984, ApJ, 278, 37
\bibitem[\protect\citeauthoryear{Errani et al.}{2015}]{Errani15} Errani, R., Penarrubia, J., Tormen, G. 2015, MNRAS, 449, L46
\bibitem[\protect\citeauthoryear{Fuller et al.}{1999}]{Fuller99} Fuller, T. M., West, M. J., Bridges, T. J., 1999, ApJ, 519, 22
\bibitem[\protect\citeauthoryear{Ganeshaiah Veena et al.}{2019}]{Veena19} Ganeshaiah Veena, P., Cautun, M., Tempel, E., van de Weygaert, R., Frenk, C. S. 2019, MNRAS, 487, 1607
\bibitem[\protect\citeauthoryear{Greco et al.}{2018}]{Greco18} Greco, J. P., et al. 2018, PASJ, 70S, 19
\bibitem[\protect\citeauthoryear{Jiang et al.}{2019}]{Jiang19} Jiang, F., Dekel, A., Freundlich, J., Romanowsky, A. J., Dutthon, A., Maccio, A., Di Cintio, A. 2019, MNRAS, 487, 5272
\bibitem[\protect\citeauthoryear{Laigle et al.}{2015}]{Laigle15} Laigle, C., et al., 2015, MNRAS, 446, 2744
\bibitem[\protect\citeauthoryear{Lee et al.}{2017}]{Lee17} Lee, M. G., Kang, J., Lee, J. H., Jang, I. S. 2017, ApJ, 844, 157
\bibitem[\protect\citeauthoryear{Liao et al.}{2019}]{Liao19} Liao, S., et al. 2019, MNRAS, 490, 5182
\bibitem[\protect\citeauthoryear{Libeskind et al.}{2013}]{Libeskind13} Libeskind, N. I., Hoffman, Y., Steinmetz, M., Gottl\"ober, S., Knebe, A., Hess, S., 2013, ApJ, 766L, 15
\bibitem[\protect\citeauthoryear{Mancera Pi\~na et al.}{2019}]{Pina19} Mancera Pi\~na, P. E., Aguerri, J. A. L., Peletier, R. F., Venhola, A., Trager, S., Choque Challapa, N. 2019, MNRAS, 485, 1036
\bibitem[\protect\citeauthoryear{Mancera Pi\~na et al.}{2020}]{Pina20} Mancera Pi\~na, P. E., et al. 2020, MNRAS, 495, 3636
\bibitem[\protect\citeauthoryear{Martin et al.}{2019}]{Martin19} Martin, G., Kaviraj, S., Laigle, C., Devriendt, J. E. G., Jackson, R. A., Peirani, S., Dubois, Y., Pichon, C., Slyz, A. 2019, MNRAS, 485, 796
\bibitem[\protect\citeauthoryear{Mart\'in-Navarro et al.}{2019}]{Martin-Navarro19} Mart\'in-Navarro, I., et al. 2019, MNRAS, 484, 3425
\bibitem[\protect\citeauthoryear{Mo et al.}{1998}]{Mo98} Mo, H. J., Mao, S., White, S. D. M., 1998, MNRAS, 295, 319
\bibitem[\protect\citeauthoryear{Niederste-Ostholt et al.}{2010}]{Niederste-Ostholt10} Niederste-Ostholt, M., Strauss, M. A., Dong, F., Koester, B. P., McKay, T. A. 2010, MNRAS, 405, 2023
\bibitem[\protect\citeauthoryear{Pahwa et al.}{2016}]{Pahwa16} Pahwa, I., et al., 2016, MNRAS, 457, 695
\bibitem[\protect\citeauthoryear{Penny \& Conselice}{2008}]{Penny08} Penny, S. J., \& Conselice, C. J. 2008, MNRAS, 383, 247
\bibitem[\protect\citeauthoryear{Pereira et al.}{2008}]{Pereira08} Pereira, M. J., Bryan, G. L., Gill, S. P. D., 2008, ApJ, 672, 825
\bibitem[\protect\citeauthoryear{Pinkney et al.}{1993}]{Pinkney93} Pinkney, J., Rhee, G., Burns, J. O., Hill, J. M., Oegerle, W., Batuski, D., Hintzen, P., 1993, ApJ, 416, 36
\bibitem[\protect\citeauthoryear{Porciani et al.}{2002}]{Porciani02} Porciani, C., Dekel, A., Hoffman, Y., 2002, MNRAS, 332 325
\bibitem[\protect\citeauthoryear{Rong et al.}{2015a}]{Rong15a} Rong, Y., Yi, S.-X., Zhang, S.-N., Tu, H. 2015a, MNRAS, 451, 2536
\bibitem[\protect\citeauthoryear{Rong et al.}{2015b}]{Rong15b} Rong, Y., Zhang, S.-N., Liao, J.-Y. 2015b, MNRAS, 453, 1577
\bibitem[\protect\citeauthoryear{Rong et al.}{2016}]{Rong16} Rong, Y., Liu, Y., Zhang, S.-N. 2016, MNRAS, 455, 2267
\bibitem[\protect\citeauthoryear{Rong et al.}{2017}]{Rong17a} Rong, Y., Guo, Q., Gao, L., Liao, S., Xie, L., Puzia, T. H., Sun, S., Pan, J. 2017, MNRAS, 470, 4231
\bibitem[\protect\citeauthoryear{Rong et al.}{2019a}]{Rong19a} Rong, Y., et al. 2019a, eprint arXiv:~1907.10079
\bibitem[\protect\citeauthoryear{Rong et al.}{2019b}]{Rong19b} Rong, Y., et al. 2019b, ApJ, 883, 56
\bibitem[\protect\citeauthoryear{Rong et al.}{2020}]{Rong20} Rong, Y., Zhu, K., Johnston, E. J., Zhang, H.-X., Cao, T., Puzia, T. H., Galaz, G. 2020, eprint arXiv:~2007.12712
\bibitem[\protect\citeauthoryear{Sales et al.}{2019}]{Sales19} Sales, L. V., Navarro, J. F., Penafiel, L., Peng, E. W., Lim, S., Hernquist, L., 2019, eprint arXiv:~1909.01347
\bibitem[\protect\citeauthoryear{Schneider et al.}{2013}]{Schneider13} Schneider, M. D., et al. 2013, MNRAS, 433, 2727
\bibitem[\protect\citeauthoryear{Scodeggio et al.}{1995}]{Scodeggio95} Scodeggio, M., Solanes, J. M., Giovanelli, R., Haynes, M. P. 1995, ApJ, 444, 41
\bibitem[\protect\citeauthoryear{Struble}{1990}]{Struble90} Struble, M. F., 1990, AJ, 99, 743
\bibitem[\protect\citeauthoryear{Tempel et al.}{2015}]{Tempel15} Tempel, E., Guo, Q., Kipper, R., Libeskind, N. I., 2015, MNRAS, 450, 2727
\bibitem[\protect\citeauthoryear{Tempel \& Libeskind}{2013}]{Tempel13a} Tempel, E., Libeskind, N. I., 2013, ApJ, 775, 42
\bibitem[\protect\citeauthoryear{Tempel et al.}{2013}]{Tempel13b} Tempel, E., Stoica, R. S., Saar, E., 2013, MNRAS, 428, 1827
\bibitem[\protect\citeauthoryear{van der Burg et al.}{2016}]{vanderBurg16} van der Burg, R. F. J., Muzzin, A., Hoekstra, H. 2016, A\&A, 590, 20
\bibitem[\protect\citeauthoryear{van Dokkum et al.}{2015}]{vanDokkum15} van Dokkum, P. G., Abraham, R., Merritt, A., Zhang, J., Geha, M., Conroy, C. 2015, ApJ, 798L, 45
\bibitem[\protect\citeauthoryear{van Dokkum et al.}{2016}]{vanDokkum16} van Dokkum, P. G., et al. 2016, ApJL, 828, 6
\bibitem[\protect\citeauthoryear{Venhola et al.}{2017}]{Venhola17} Venhola, A., et al. 2017, A\&A, 608, 142
\bibitem[\protect\citeauthoryear{West et al.}{2017}]{West17} West, M. J., de Propris, R., Bremer, M. N., Phillipps, S., 2017, Nature Astronomy, 1, 157
\bibitem[\protect\citeauthoryear{Yagi et al.}{2016}]{Yagi16} Yagi, M., Koda, J., Komiyama, Y., Yamanoi, H. 2016, ApJS, 225, 11
\bibitem[\protect\citeauthoryear{Yozin \& Bekki}{2015}]{Yozin15} Yozin, C., Bekki, K. 2015, MNRAS, 452, 937
\bibitem[\protect\citeauthoryear{Zhang et al.}{2009}]{Zhang09} Zhang, Y., Yang, X., Faltenbacher, A., Springel, V., Lin, W., Wang, H., 2009, ApJ, 706, 747
\bibitem[\protect\citeauthoryear{Zhang et al.}{2015}]{Zhang15} Zhang, Y., Yang, X., Wang, H., Wang, L., Luo, W., Mo, H. J., van den Bosch, Frank C., 2015, ApJ, 798, 17

\end{thebibliography}


\end{document}